\documentclass[twocolumn,amsmath,amssymb,floatfix,prd]{revtex4}

\usepackage{bm}
\usepackage{amsmath}
\usepackage{epsf}

\newcommand{\beq}{\begin{equation}}

\newcommand{\eeq}{\end{equation}}
\newcommand{\beqa}{\begin{eqnarray}}
\newcommand{\eeqa}{\end{eqnarray}}

\newcommand{\lagr}{{\cal L}}
\newcommand{\zero}{{\rm min}}

\def\simlt{\lesssim}
\def\simgt{\gtrsim}

\newcommand{\ApJS}{Astrophys. J Supp.}

\newcommand{\ApJ}{Astrophys. J}

\newcommand{\mnras}{Mon. Not. R. Astron. Soc.}

\begin{document}

\title{Challenges for Kinetic Unified Dark Matter}
\author{Dimitrios Giannakis$^{1}$ and Wayne Hu$^{2}$}
\affiliation{
$^{1}$Department of Physics,
University of Chicago, Chicago IL 60637 \\
$^{2}$Department of Astronomy and Astrophysics,
University of Chicago, Chicago IL 60637
}

\begin{abstract}
\baselineskip 11pt Given that the dark matter and dark energy in the
universe affect cosmological observables only gravitationally, their
phenomenology may be described by a single stress energy tensor.  
True unification however  requires a theory that reproduces the
successful phenomenology of $\Lambda$CDM and that requirement
places specific constraints on the  stress structure of the matter.
We show that
 a recently proposed unification through an offset
quadratic kinetic term for a scalar field is exactly equivalent to  a
fluid with a closed-form barotropic equation of state plus
cosmological constant.  The
finite pressure at high densities introduces a cutoff in the linear
power spectrum, which may alleviate the dark matter substructure
problem; we provide a convenient fitting function for such studies.
Given that sufficient power must remain to reionize the universe, the
equation of state today is nonrelativistic
with $p \propto \rho^{2}$ and a Jeans scale in the parsec
regime for all relevant densities.  Structure may then be evolved
into the nonlinear regime with standard hydrodynamic techniques.  In
fact, the model is equivalent to the well-studied collisional dark
matter with negligible mean free path.  If recent observations of the
triaxiality of dark matter halos and ram
pressure stripping in galaxy clusters are confirmed, this model will
be ruled out.
\end{abstract}
\maketitle

\section{Introduction}
\label{sec:introduction}

The standard model for cosmology contains dark energy in the form of
a cosmological constant and cold dark matter ($\Lambda$CDM).  While
this model explains most observations to date, the dark matter and
the dark energy are only observed gravitationally. It thus remains
possible that they are a single phenomenon, sometimes called in the literature
unified dark matter (UDM).  

A candidate for UDM that has been extensively discussed in the literature
is the generalized  Chaplygin gas. Here the UDM is taken to be a perfect 
fluid with an equation of state $p =-A/\rho^\alpha$, with $A > 0$ and 
$0 < \alpha \leq 1$; models with $\alpha=1$  \cite{KamMosPas01,BilTupVio02}
are particularly attractive since they represent a Born-Infeld
scalar field \cite{BenBerSen02,PadCho02} that admits a string
theory interpretation  (e.g.~\cite{Sen02,Gib98}).  This class of equations
of state leads to a {\it background} evolution that transitions from a 
matter dominated phase at early times to a de Sitter phase in the 
asymptotic future in agreement with distance measures 
\cite{MakOliWag03,BerSenSenSil04,BenBerSen03}. 

Unfortunately,  in linear perturbation
theory Chaplygin UDM produces oscillations or exponential growth
in the matter power spectrum unless $\alpha \rightarrow 0$ 
\cite{SanTegZalWag04}, even after the
effect of baryons \cite{BecAveCarMar03} has been included. The exact phenomenology
of these models once nonlinear structures have formed is complicated due to the
fact that the pressure decreases with increasing energy density
(see e.g. \cite{Aveetal04,Biletal04}).  It is clear,
however, that these models differ substantially from the successful
$\Lambda$CDM model and are unlikely to be viable in their simplest form.

These problems with Chaplygin gas UDM stem from
very general considerations and serve as a guide to building more 
successful UDM models.
Fundamentally, a UDM
model that mimics the joint stress energy tensor of $\Lambda$CDM
will reproduce all of its observational successes \cite{HuEis99}.
Since
the UDM stress-energy tensor is covariantly conserved, mimicry in
the spatial stresses is all that need be explicitly enforced
\cite{Hu98}.

Deviations from perfect mimicry must be small in a very specific sense.  
The basic requirement for mimicry is that spatial fluctuations in
the UDM stresses be negligible in comparison to those in its energy
density $\delta \rho$, even though the homogeneous or background
stress must be of order $-\rho$ to accelerate the expansion
\cite{HuEis99,SanTegZalWag04}. More specifically, the ratio between
the fluctuating stress and the energy density must be of order the
{\it square} of that between the momentum and energy densities,
 or a non-relativistic  velocity squared.   This disparity
implies that the total pressure is effectively decoupled from the
total density, i.e.~that the pressure is effectively non-adiabatic
\cite{SanTegZalWag04,ReiWagCalJor03,ReiMakWag05}.  For
a generalized Chaplygin gas this requirement is only satisfied as $\alpha
\rightarrow 0$.

In the linear regime, where the momentum density itself is a
perturbation, fluctuations in the spatial stresses must then be a
second order effect quantified by a vanishingly small sound speed
and anisotropic stress. The internal dynamics of dark matter halos
place even more
stringent requirements on mimicry.  Here the UDM stress energy tensor 
must embed the information
from multiple momentum streams in the dark matter, i.e. the full
phase space structure of cold dark matter.   A successful UDM model
will be at least as complex as $\Lambda$CDM in order to reproduce
its rich phenomenology detracting somewhat from its appeal.

The prototype for a successful UDM model is a scalar field coherently
oscillating in an offset quadratic potential \cite{HuEis99}; it
satisfies mimicry requirements all the way down to the de-Broglie
wavelength of the field \cite{HuBarGru00}.  It does not, however,
explain the relationship between the $10^{-33}$ eV flatness of the
offset scale and the $>10^{-22}$ eV mass scale.



Recently, Scherrer \cite{Sch04} introduced a $k$-essence
\cite{ArmMukSte00} variant of scalar field UDM that also succeeds
in linear theory.  Here the quadratic
potential is replaced by a quadratic kinetic term (see
\cite{ArkCheLutMuk03} for a possible physical motivation). An important
feature of this model is that its stress-energy tensor is purely kinetic
(see \cite{ChiOkaYam00} for a generalization).  
We shall show that its equation of state is consequently both
barotropic and expressible in closed form.  We are consequently able
to reduce this kinetic form of UDM to 
collisional dark matter with a negligible mean free path.
Non-linear structure formation for this case has been well-studied in the literature
\cite{Mooetal00,YosSprWhi00}.

The outline of the paper is as follows. We begin in \S
\ref{sec:scalar} with a brief review of the basic properties of scalar field UDM.  We
derive an exact, fully non-linear, equation of state for the purely kinetic UDM model \cite{Sch04} in 
\S \ref{sec:eos}. We
then calculate the evolution of linear density fluctuations in \S
\ref{sec:linear}.   We place constraints on the model from the suppression of the low mass
dark matter halo abundance in \S \ref{sec:nonlinear1} and discuss opportunities and challenges
for the model from halo substructure in \S \ref{sec:nonlinear2}. 
We conclude in \S \ref{sec:discussion}. 

Throughout we will compare the UDM model to a $\Lambda$CDM  model
with a  total non-relativistic matter density
$\Omega_{m}h^{2}=\Omega_{c}h^{2}+\Omega_{b}h^{2}=0.14$, baryon
density $\Omega_{b}h^{2}=0.024$, in a flat cosmology with
$\Omega_{\Lambda}=0.73$ and scale invariant adiabatic initial
conditions.


\section{Scalar Field UDM}
\label{sec:scalar}

In this section, we briefly review the relationship 
between scalar fields and perfect fluids and
show why scalar fields whose potential and kinetic terms possess a
non-vanishing minimum value satisfy the basic 
requirements of a successful UDM model. 
The minimum acts as a decoupled source of 
negative pressure with an equation of state $p/\rho=-1$ for the
background that does not contribute to stress fluctuations. 

 The stress energy tensor of a classical scalar field $Q$ with a Lagrangian $\lagr$ and kinetic
 term
 \begin{equation}
 X = -{1\over 2}\nabla_\mu Q \nabla^\mu Q
 \end{equation}
 is given by
\begin{align}
{T^\mu_{\hphantom{0}\nu}} =  \lagr_{,X} \nabla^\mu Q \nabla_\nu Q   + \lagr \delta^\mu_{\hphantom{i}\nu} \,.
\label{eqn:scalarfieldt}
\end{align}
It takes the form of a perfect fluid
\begin{align}
{T^\mu_{\hphantom{0}\nu}} =  (\rho + p) U^\mu U_\nu + p \delta^\mu_{\hphantom{i}\nu}\,,
\label{eqn:Tmunu}
\end{align}
with the association of a 4-velocity \cite{GarMuk99}
\begin{align}
U_\mu = {\nabla_\mu Q \over (2 X)^{1/2}} \,,
\label{eqn:fourvelocity}
\end{align}
which is timelike for the nearly homogeneous cosmological initial
conditions. We shall show in \S \ref{sec:nonlinear2} that it remains
so throughout structure formation. Here, the proper energy density
and pressure, or energy density and isotropic stress as measured by
an observer comoving with the ``fluid", is
\begin{align}
\rho &= 2 X \lagr_{,X} - \lagr\,, \qquad p = \lagr\,.
\end{align}
Although the mapping from a scalar field to a perfect fluid is exact, the fluid equations
are not closed without a specification of $\lagr$ to provide the equation of
state.

Scalar fields provide fertile ground for unified dark matter models.
When combined, the kinetic and potential terms possess
sufficient degrees of freedom to alter the equation of state of the fluctuations
separately from the background.
  Both the potential and the kinetic versions of UDM
posit a separable Lagrangian of the form
\begin{equation}
\lagr = \lagr_e + \lagr_m \,,
\end{equation}
where $\lagr_e$ and $\lagr_m$ behave respectively as dark energy and dark matter.
Mimicry of a cosmological constant is possible if $\lagr_e$ is
constant or, more generally, slowly varying with the field degrees
of freedom in comparison to the matter piece $\lagr_m$. In what
follows, we shall assume that the former case holds and set
\begin{equation}
 -\lagr_e = \rho_\Lambda = {3 H_0^2 \Omega_\Lambda \over 8\pi G}\,.
 \end{equation}
 The stress energy tensors $\lagr_e$ and $\lagr_m$ are then separately
 conserved. Other splittings are of course possible
 (e.g.~$\rho_e = - \lagr$ \cite{PadCho02}) but all choices yield the
 same observable predictions.  Our choice simplifies the calculation in that
 the dark matter and dark energy pieces interact only through gravity.

 The prototype for this sort of unification is an axion-like matter component
\begin{equation}
\quad \lagr_m =  X-m^2 (Q -Q_\zero)^2
\end{equation}
  in a potential whose minimum gives the dark energy density $\lagr_e = -V(Q_\zero)$.
 We will refer to this option as potential UDM (pUDM).

 Scherrer \cite{Sch04} introduced a similar but alternate ansatz based on a modified kinetic term
\begin{equation}
\lagr_m = F (X-X_\zero)^2\,;
\end{equation}
we call this kinetic UDM (kUDM).
The constant $F$ can be eliminated in favor of the average matter density and kinetic term
\begin{align}
\bar\rho_m &= F X_\zero^2\epsilon ( 4 + 3\epsilon ) \,, \quad \bar\rho_{m\,0} =
  {3 H_0^2 \Omega_c \over 8\pi G}\,,
\end{align}
where $\epsilon = (\bar X - X_\zero)/X_\zero$ and $``0"$ denotes evaluation at
the present epoch.

The fine tuning between the mass and offset scale in pUDM discussed in \S \ref{sec:introduction}
still exists in kUDM as the coincidence problem $\bar\rho_{m \,0} \sim
\rho_\Lambda$.  We shall see that  kUDM also requires $\epsilon_0 \simlt
10^{-18}$ (see \S \ref{sec:nonlinear1}), which is
similar to the pUDM requirement of a purely quadratic potential near
the minimum or the neglect of self interaction terms in the
potential.  

Nonetheless, both the pUDM and kUDM model satisfy the basic requirement for a successful UDM model
in that their Lagrangian reaches a non-vanishing minimum value $\lagr_e$ that can
serve as a dark energy component that cannot be spatially perturbed.

\section{Equation of State}
\label{sec:eos}

Given the mapping between scalar fields and perfect fluids described 
in the previous section,
the dynamics of the kUDM model are determined by its equation of state, the
relationship between its pressure and energy density in its rest frame.  
An important property of this purely kinetic model is that there is only one
degree of freedom, the kinetic term $X$, and therefore the equation of
state is barotropic 
to all orders in the density fluctuation, i.e. $p_m$ is a function of $\rho_m$ only.

We begin with the background equation of state
\begin{equation}
w_m(a) \equiv {\bar p_{m} \over \bar \rho_{m}} = {\epsilon(a) \over 4 + 3\epsilon(a)}\,.
\end{equation}
To obtain $\epsilon(a)$ note that energy conservation provides
an explicit relation for 
\begin{equation}
a(\epsilon) = \left[ {\epsilon^{2} (1+\epsilon) \over
\epsilon_{0}^{2}(1+\epsilon_{0} )}\right]^{-1/6}\,.
\end{equation}

A convenient fitting formula for
the required inverse relation
\begin{equation}
\epsilon(a) \approx \epsilon_0 a^{-3} \left\{ 1 + \left[  \left( 1+\epsilon_0^{-1} \right)^{1/3} a \right]^{-\gamma}
\right\}^{-1/\gamma}\,,
\end{equation}
achieves $\sim 1\%$ accuracy
with $\gamma=5/2$. Thus, the ``matter" component behaves as
radiation in the background for $a \ll \epsilon_{0}^{1/3}$ and
non-relativistic dark matter thereafter.

Since the matter component depends only on $X$, the pressure is
adiabatic and the fully non-linear equation of state
\begin{align}
p_m(\rho_m) &= \rho_m \left[ {1 \over 3} + {2 \over 3 f} (1 - \sqrt{1+f}) \right]\,, \nonumber\\
f &= {3 \rho_m\over 4\bar\rho_m }\epsilon (4+3\epsilon)\,,
\label{eqn:eos}
\end{align}
is barotropic.  The adiabatic sound speed
\begin{align}
{dp_m \over d\rho_m} = {p_{m,X} \over \rho_{m,X}} = {1 \over 3} - {1 \over 3} {1 \over \sqrt{1+f}} \,,
\label{eqn:sound}
\end{align}
then quantifies the pressure response to a density fluctuation in
both the linear and non-linear regimes.
Note that the pressure is a monotonically increasing function of the energy density.
Hence, unlike the Chaplygin gas model \cite{Aveetal04,Biletal04}, 
the matter cannot be destabilized  by the appearance of
high density fluctuations.   We will exploit this fact in \S \ref{sec:nonlinear2} in
calculating the abundance of collapsed objects. 

Note that the simplification of a purely kinetic UDM field is an important feature
in the model that is critical in the sections that follow.
  In the more general case
(including pUDM), where $\lagr$ is a function of both $Q$ and $X$,
the pressure fluctuations $p(X,Q)$ cannot be expressed as a function
of $\rho$, except in the background. In particular, they may be
highly time variable.  Regardless, the scalar field convention is to
define the effective sound speed as that of the kinetic degree of
freedom \cite{Hu98,GarMuk99}, i.e.
\begin{equation}
c_e^2 \equiv {\delta p \over \delta \rho} \Big|_{U_i=0} = {p_{,X} \over \rho_{,X}} \,,
\end{equation}
since it quantifies  the stresses in the frame comoving with the
field. However, this mapping of field variables onto fluid variables
is merely formal in the general case.  For example, the matter
component of pUDM acts as collisionless non-relativistic dark matter
even though $c_e^2=1$
 (see \S
\ref{sec:discussion}).  In the kUDM model, no such subtlety exists and
the matter component behaves purely hydrodynamically. 

\begin{figure}[tb]
\centerline{\epsfxsize=3.4in\epsffile{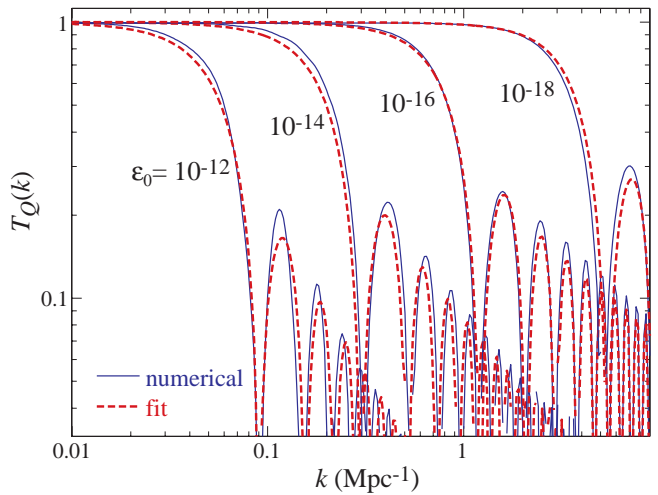}}
\caption{Modification to the matter transfer function introduced by
kUDM as a function of $\epsilon_0$, which quantifies the deviation
of $\bar X$ from $X_\zero$ today.  The fluctuations oscillate
acoustically and decay at small scales in a manner that can be
conveniently fit by Eqn.~(\ref{eqn:fit}).} \label{fig:trans}
\end{figure}

\section{Linear Regime}
\label{sec:linear}

For the linear fluctuations it is sufficient to note that though
there is a single kUDM Lagrangian, the matter field component is
decoupled from the dark energy component and behaves as a fluid.  It possesses
an adiabatic sound speed given by Eqn.~(\ref{eqn:sound}) with $f=
3\epsilon(1+3\epsilon/4)$ and negligible anisotropic stress, as the
latter scales quadratically with spatial field derivatives [see
Eqn.~(\ref{eqn:Tmunu})].

Given that the only gravitational degrees of freedom involve the
parametrization of the stresses, mimicry of $\Lambda$CDM for all
linear theory observables above the sound horizon, including the cosmic microwave
background (CMB)
and large-scale structure, is guaranteed by construction
 \cite{HuEis99}.  Perfect mimicry follows from  $\lagr_e$ being strictly
 constant,
since the spatial fluctuations $\delta \rho_\Lambda$ and $\delta
p_\Lambda$ vanish independently of the scalar field sound speed.
More generally, allowing $\lagr_e$ to depend weakly
on $Q$ or $X$ can modify the predictions on the largest scales by
introducing a relativistic dark energy sound speed
\cite{CalDavSte98,Hu98,BeaDor04}.

Below the sound horizon, perturbations in the matter component
undergo Jeans or acoustic oscillations instead of growth,
effectively cutting off the linear power spectrum.   This behavior
places an upper limit on $\epsilon_{0}$ such that the matter
component acts as non-relativistic matter during the usual matter
dominated epoch $a>a_{\rm eq}=3\times 10^{-4}(\Omega_m
h^2/0.14)^{-1}$.  Scherrer placed a conservative upper limit by
demanding that $\epsilon(a_{\rm eq}) < 1$ or, equivalently,
$\epsilon_0 <  4 \times 10^{-11} ( \Omega_m h^2/0.14)^{-3}$
\cite{Sch04}.  This constraint would ensure that the kUDM
 behaves as $\Lambda$CDM above the horizon at matter radiation
equality and thus only deep in the linear regime today.   To ensure
mimicry for non-linear structure requires a much more stringent
constraint on $\epsilon_0$.

These considerations can be usefully quantified by noting that the
 equation of state $w_m$ and sound speed drop rapidly and monotonically
  from their relativistic values
$w_{m}= dp_{m}/d\rho_{m}=1/3$ once $a > a(\epsilon=1)$.  After a
scale exits the sound horizon, matter fluctuations grow as CDM
independently of scale.  Thus, on all larger scales, linear
fluctuations can be described by a modification to the usual
transfer function (e.g. \cite{EisHu99})
\begin{equation}
T(k) = T_{\rm CDM}(k) T_{Q}(k) \,.
\end{equation}
To characterize $T_Q$, we solve the coupled Einstein-Boltzmann
equations that include photons, baryons and neutrinos in the usual
way, following \cite{Hu98,GorHu04}. The resulting transfer function,
shown in Fig.~\ref{fig:trans}, is well fit by
\begin{align}
T_Q(k) &\approx {3 j_1(x) \over x}\left[1+\left( {x \over 3.4}\right)^2\right]^{1/(\beta+1)} \,,
\label{eqn:fit}
\end{align}
with
\begin{align}
x &= \left( k \eta_* \over 7.74 \right)\,, \quad
\beta = 0.21 \left[ {\epsilon_0 \over 10^{-18} } \left( {\Omega_m h^2 \over 0.14} \right)^3
\right]^{0.12} .
\end{align}
Here, $\eta_*$ is the conformal time evaluated at $a_* = 14
\epsilon_0^{1/3}$. Note that the power drops by at least an order of
magnitude at $x=5.9$, the first extremum of the oscillation.    For
$\epsilon_0 \rightarrow 0$, this drop reflects the lack of the usual
logarithmic growth of the matter fluctuations in the radiation
dominated era.  For larger $\epsilon_0$, the fluctuations are
further suppressed as a function of $k$ due to the change in the
growth rate in the matter dominated epoch.

\begin{figure}[tb]
\centerline{\epsfxsize=3.4in\epsffile{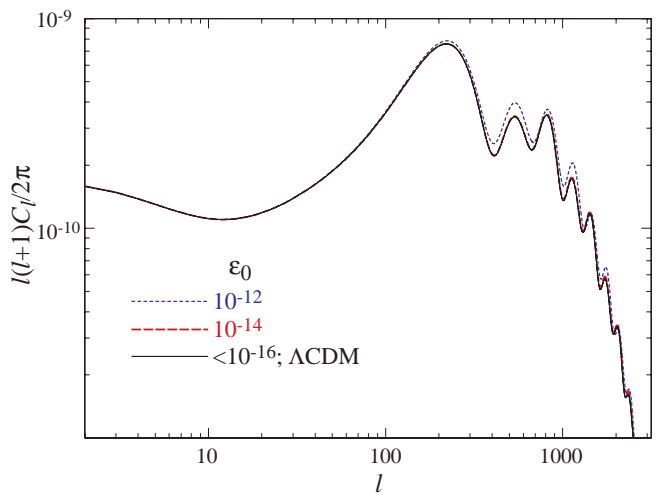}}
\caption{CMB power spectrum in the fiducial cosmological model as a function of $\epsilon_0$.
For the small $\epsilon_0< 10^{-16}$ required to match early structure, CMB predictions are
indistinguishable from $\Lambda$CDM.}
\label{fig:cl}
\end{figure}

In Fig.~\ref{fig:cl} we show that the CMB power spectrum predictions
remain unchanged from $\Lambda$CDM to the cosmic variance limit for
$\ell \simlt 10^3$ and $\epsilon_0 \simlt 10^{-16}$. The main effect
of higher $\epsilon_0$ is an enhancement of the radiation driving
and a reduction of the baryonic modulation of the peaks.

\section{Halo Abundance}
\label{sec:nonlinear1}

Given that the background pressure of the matter component of kUDM
drops as the universe expands, the largest qualitative difference
between this model and $\Lambda$CDM in the non-linear regime comes
simply from the suppression of small scale linear power that is
frozen in early on.

\begin{figure}[tb]
\centerline{\epsfxsize=3.4in\epsffile{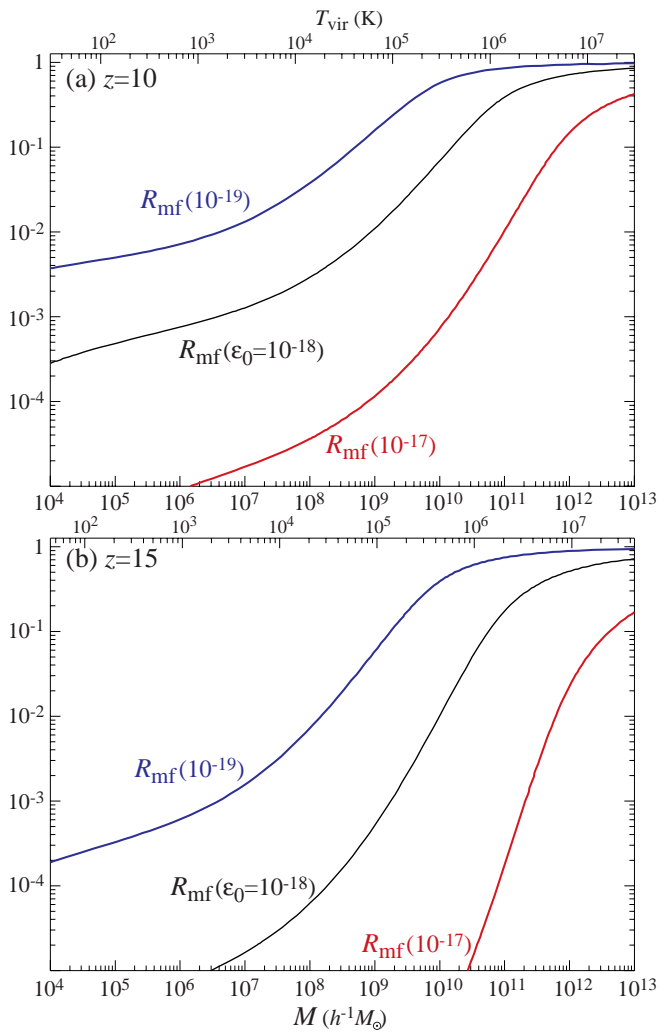}}
\caption{Ratio of mass functions $R_{\rm mf}$
 in kUDM and $\Lambda$CDM.  The comoving number density
of halos of mass $M$ or virial temperatures of $T_{\rm vir}$ is suppressed due to the
cut off in the power spectrum from linear theory.   The suppression increases with decreasing
mass and increasing redshift.} \label{fig:massfn}
\end{figure}

This suppression in the initial fluctuations
reduces the number of low mass dark matter halos
in a manner that has been well quantified by simulations.
For Gaussian intial conditions, 
the number density of halos of mass $M$ is controlled by the variance of linear density
field \cite{PreSch74}
\begin{equation}
\sigma^2(M;z,\epsilon_0) = \int\frac{d^3k}{(2\pi)^3} P_{\rm CDM}(k;z) T_Q^2(k;\epsilon_0)|W(k;M)|^2.
\end{equation}
where $W(k;M)$ is the Fourier transform of the top hat window that encloses the mass $M$ at
the mean density.
Specifically, their comoving number density  is described
by the mass function \cite{SheTor99}
\begin{eqnarray}
{d n \over d\ln M}= {\bar \rho_{m\,0} \over M} f(\nu) {d\nu \over d\ln M}\,,
\label{eqn:massfn}
\end{eqnarray}
where 
$\nu = \delta_c/\sigma(M)$ and
\begin{eqnarray}
\nu f(\nu) = A\sqrt{{2 \over \pi} a\nu^2 } [1+(a\nu^2)^{-p}] \exp[-a\nu^2/2]\,.
\label{eqn:stform}
\end{eqnarray} 
A choice of 
$\delta_c=1.69$, $a=0.75$, $p=0.3$, and $A$ such that $\int d\nu f(\nu)=1$
fits the results of simulations well.

At high redshift, where $\sigma$ is small and $\nu$ large, 
collapsed objects of a given mass
are rare.  In this regime, the number density is exponentially sensitive to $\sigma(M;z,\epsilon_{0})$
and hence the amplitude of initial fluctuations.  Thus, the requirement that there
are sufficient low mass halos at high redshift to reionize the universe places
a constraint on kUDM models.

Let us define the suppression in kUDM for a given $\epsilon_{0}$
relative to $\Lambda$CDM in the mass function
as
\begin{equation}
R_{\rm mf}(\epsilon_{0}) \equiv {dn/d\ln M\, (\epsilon_{0}) \over dn /d\ln M\, (0)}\,.
\end{equation}
In  Fig.~\ref{fig:massfn} we show this suppression as a function of mass
for two  redshifts representative of reionization and several choices of $\epsilon_{0}$.
Note that the suppression increases with decreasing mass due to the small-scale
cut off in the transfer function $T_{Q}$ and with increasing redshift due to the 
exponential sensitivity.

\begin{figure}[tb]
\centerline{\epsfxsize=3.4in\epsffile{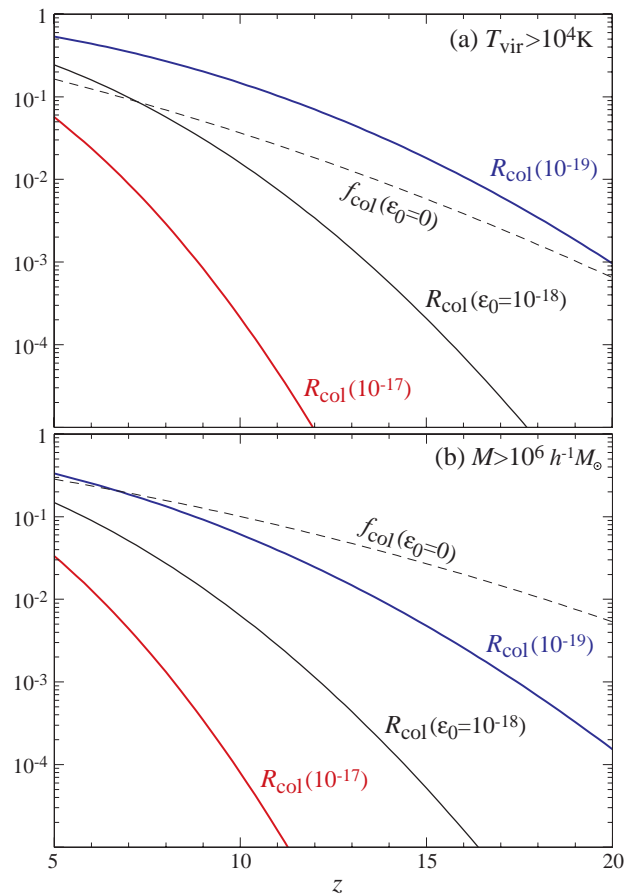}}
\caption{Ratio of collapse fractions $R_{\rm col}$ between
kUDM and $\Lambda$CDM
in halos of mass relevant for reionization.  Top panel: halos that can cool by atomic line transitions
$T_{\rm vir}> 10^{4}$K.  Bottom panel: by molecular hydrogen cooling $M \simgt 10^{6} h^{-1}M_{\odot}$.  Also shown for reference is the collapse fraction itself for $\Lambda$CDM
$f_{\rm col}(0)$.} \label{fig:collapse}
\end{figure}

For the specific question of reionization, the relevant quantity for
comparison
is the fraction of the universe that has collapsed into objects of sufficient mass that 
they can cool and form stars.  
  For atomic line cooling, the virial temperature of the
halo
\begin{align}
T_{\rm vir} = 1.3 \times 10^4 {\rm K} \left( {\Omega_m \over 0.27} \right)^{1/3} \left( {{1+z} \over 10}\right)
\left( { M \over 10^8 h^{-1} M_\odot} \right)^{2/3}
\end{align}
must be greater than $\sim 10^{4}$K; for molecular hydrogen cooling, $M\simgt 10^{6} h^{-1} M_\odot$
but the reionization efficiency is lower
\cite{BarHaiOst01,YosAbeHerSug03,SomBulLiv03}.  The collapse fraction is given
by the integral over the mass function
\begin{align}
f_{\rm col} = \int d\ln M  {M \over \bar \rho_{m\,0}} {dn \over d\ln M} \,.
\end{align}
We will again define the collapse fraction relative to $\Lambda$CDM as
\begin{align}
R_{\rm col}(\epsilon_{0}) \equiv {f_{\rm col}(\epsilon_{0}) \over f_{\rm col}(0)}\,.
\end{align}
We show this suppression factor in Fig.~\ref{fig:collapse}.

Since even $\Lambda$CDM has difficulty reionizing the universe sufficiently early,
this ratio cannot be smaller than $R_{\rm col} \sim 10^{-2}$ around $z\sim 10$.  Comparing requirements in the very similar
context of 
broken scale invariance and warm dark matter
\cite{BarHaiOst01, NarSpeDavMa00,SomBulLiv03} we place a conservative bound of
 $\epsilon_0 \simlt 10^{-18}$ for atomic cooling.  If reionization occurs through molecular hydrogen
 cooling, then the limit is even more stringent.  Furthermore if indications from the CMB
polarization measured by WMAP of a high reionization redshift $z>10$
\cite{Kogetal03} are confirmed then even this regime will
be 
ruled out, given reasonable assumptions on star formation and the
escape fraction of ionizing photons \cite{SomBulLiv03}.  
The only loophole is if scale invariance in the initial power spectrum is strongly
broken at coincidentally similar scales to compensate.

\section{Halo Substructure}
\label{sec:nonlinear2}

The kUDM model differs from $\Lambda$CDM in the structure of
dark matter halos both because of the reduction in initial small 
scale power and the fact that it behaves as a nearly pressureless fluid 
instead of a set of collisionless particles.  These issues will require full 
cosmological simulations to resolve quantitatively,  which is beyond
the scope of this work.  However, we shall see that both aspects of the 
modifications can be reduced to cases that have been previously 
solved in the literature.  The latter is in qualitative disagreement 
with some recent observations.
 
A suppression of small scale power may in fact be desirable to reduce the claimed excess
present-day substructure in galactic sized halos of the $\Lambda$CDM model.
Of course, there is tension between these beneficial effects and the inability to
form enough high redshift halos to reionize the universe.   Saturating the conservative
bound by setting $\epsilon_0 \sim 10^{-18}$ one recovers a suppression of
small scale power at a similar scale and form to the well-studied broken
scale invariance model \cite{KamLid00,ZenBul03}.  This model was explicitly
constructed to alleviate the substructure problem and so there may remain a small
window in kUDM parameter space that satisfies these conflicting requirements.
 Cosmological simulations will be required to address this
question quantitatively.  Note, however, that the substructure
problem may have an astrophysical origin.

For the evolution of small scale structure, it is useful to bear in mind that
Eqn.~({\ref{eqn:eos}}) implies that  the pressure remains
small compared with the energy density until $f=1$. Given the low
upper bound for $\epsilon_{0}$, this corresponds to extremely large
overdensities today $\rho_{m\,0}/\bar \rho_{m\,0} \sim
1/\epsilon_{0} \sim 10^{18}$.  For the relevant
 $f\ll 1$ case,  the non-linear equation of state simplifies to 
\begin{equation}
p_m = {\rho_m^2 \over 4\bar\rho_m}\epsilon
\end{equation}
and the sound speed to
\begin{equation}
{d p_m \over d\rho_m } = {1 \over 2}{\rho_m \over \bar \rho_m} \epsilon\,.
\end{equation}
Note that in this limit the fluctuation to the kinetic term is still
small, since
\begin{equation}
{\delta X \over X_\zero} = \epsilon {\delta\rho_m \over \bar\rho_m} \ll
1.
\end{equation}
Hence, the fluid 4-velocity [see Eqn.~(\ref{eqn:fourvelocity})]
remains timelike (see~\cite{ArmPicLim05} for a discussion of
spacelike kinetic scalar fields) and the 3-velocity
non-relativistic. Thus, the nonlinear evolution for kUDM may be
calculated with non-relativistic hydrodynamic techniques, such as
smoothed particle hydrodynamics (e.g.~\cite{SprYosWhi01,SprHer02}).

Qualitatively, density fluctuations will stabilize below the Jeans scale, which is the
distance sound can travel in a dynamical time.  The comoving Jeans wavenumber
becomes
\begin{align}
\label{eq:kJ} k_J &= \sqrt {4\pi G\rho_m a^2 \over {dp_m / d\rho_m}
}
= \sqrt{ 8\pi G \bar \rho_m a^2 \over \epsilon} \nonumber\\
   & = 0.216 a \left( {\epsilon_0 \over 10^{-18}  } \right)^{-1/2} \left( {\Omega_m h^2 \over 0.14} \right)^{1/2}
{\rm pc}^{-1}\,,
\end{align}
or a physical Jeans scale that is fixed in the matter dominated epoch for both linear and non-linear
density fluctuations.
For a model that has sufficient linear theory power to explain high redshift structure
and reionization, this scale is well below cosmologically interesting scales.

Given these constraints, the matter component of kUDM acts as a
pressureless fluid on cosmologically relevant scales. This type of
dark matter has been studied extensively in the literature in the
context of collisional or self-interacting dark matter with a
vanishingly small mean free path \cite{Mooetal00,YosSprWhi00}. Qualitatively, cold dark matter
behaves differently from fluid dark matter as soon as there exist
multiple streams in the dark matter.  Since cold dark matter is
collisionless, it occupies a phase space distribution
 $f({\bf x},{\bf q},\eta)$, defined in a locally orthonormal basis.
 Its gravitational influence comes through the stress energy
 tensor
\begin{align}
{T^\mu_{\hphantom{0}\nu}}  = g\int{ d^{3}q \over (2\pi)^{3}} {q^{\mu} q_{\nu} \over E(q)} f\,,
\end{align}
where $q^{\mu}$ is the 4-momentum of the CDM particles and $g$ is
the degeneracy factor. For CDM in the non-linear regime, there can
be multiple streams or momentum states occupied for a given position
state. On the other hand, kUDM trajectories cannot cross and instead
kUDM will behave as a fluid and shock.

Specifically, kUDM or fluid dark matter is challenged by
observations of a galaxy cluster system observed to be undergoing a
high velocity merger \cite{Maretal04}. Here, the collisional
baryonic gas  lags behind the collisionless stars and
lensing-observed dark matter due to ram pressure stripping.
Likewise, kUDM is in potential conflict with the shape of
gravitational wakes of galactic halos moving in a cluster halo
\cite{FurLoe02} and the fundamental plane of ellipticals in clusters
versus the field \cite{GneOst01}. Finally, kUDM would not predict as
large a triaxiality of halos as measured by strong lensing
\cite{Mir02,Mooetal00,YosSprWhi00,NatLoeKneSma02}, although some ambiguities remain
due to projection effects \cite{SanTreEll02}. Note that the kUDM is
fluid-like even down to scales where the pressure becomes relevant
so that there is no equivalent to an adjustable mean free path or
cross section to alleviate these difficulties \cite{SpeSte00}.

\section{Discussion}
\label{sec:discussion}

Building a successful unified dark matter model is a useful exercise,
even if its end role is only as a foil to $\Lambda$CDM that highlights 
the successes of the standard model.  Observations already sharply
delineate the basic properties of a successful unified model.

Compared with Chaplygin gas models of unified dark matter, 
kinetic unified dark (kUDM) \cite{Sch04} is more successful as
it can explain the well-established 
observations both of the background expansion rate and phenomena in the linear
regime, such as the cosmic microwave background and large-scale structure of the
universe.  It is not without its problems however in the non-linear regime.
If recent observations in clusters of galaxies are confirmed then even this type of
model will be ruled out.

We have shown that the kUDM behaves as a fluid
with a closed-form barotropic equation of state plus a constant term that acts as a cosmological
constant to all orders in structure formation.
In linear theory, the high density of the matter component in the early universe introduces a relativistic pressure and hence a cut off to
the linear power spectrum from Jeans oscillations.   We have provided a convenient fitting
function to the numerical results that should assist in future investigations of the model.

 In order to produce halos that reionize the universe,
kUDM must behave like CDM in linear theory on scales above $k\sim$ few
Mpc$^{-1}$.  This requirement  reduces the viable kUDM parameter space to the limiting
case of highly collisional, or self-interacting, non-relativistic
dark matter with a negligible mean free path
\cite{Mooetal00,YosSprWhi00}. kUDM is then also phenomenologically
equivalent to a Chaplygin gas UDM, where the pressure is independent
of the density \cite{SanTegZalWag04} ($\alpha \rightarrow 0$, see \S \ref{sec:introduction}). 
Such models are in moderate
observational conflict with the observed shape of cluster halos and
effects like ram pressure stripping in cluster substructure.
Hydrodynamic simulations of the dark matter with the initial
conditions provided by the linear theory given here may be used to
quantitatively address these issues in the specific kUDM context.
Note that these qualitative features are likely to remain valid even
if the quadratic assumption for the kinetic Lagrangian fails far
away from the minimum, so long as the implied pressure is still of
order the energy density once $\epsilon\simgt 1$.

If these recent observations of clusters of galaxies are confirmed
only the potential variant of scalar field UDM remains viable.  Here
although the field can
still be mapped onto a perfect fluid, it possesses a phase space
structure and mimics CDM down to the de-Broglie wavelength of the
field, i.e.~the kiloparsec regime for $m > 10^{-22}$ eV
\cite{HuBarGru00}.  In this case, the rapidly oscillating equation
of state makes a direct solution of the fluid equations intractable
\cite{KodSas84,KhlMalZel85}.  It can be shown through a WKB
approximation that the momentum information is carried by a
spatially varying phase to the field, similar to a quantum phase
space for the wavefunction \cite{WidKai93}.   The collisionless
nature of CDM is then reflected by the linearity of the wave
equation or the quadratic nature of the potential. Self-interactions
for the pUDM model are included by adding anharmonic terms to the
potential \cite{Pee00}.

From these examples, it is clear that a successful UDM model will
likely be as complex phenomenologically as the well-tested
$\Lambda$CDM model.   Complexity reduces the appeal of the UDM
hypothesis, unless a fundamental theory can explain the disparate
scales introduced by the components that masquerade as the dark
matter and dark energy.  Neither the kUDM or pUDM models in the form
presented here do so.

{\it Acknowledgments:} We  thank C. Armendariz-Picon,
Y. Ascasibar, S. Carroll, C. Gordon, A. Kravtsov, R.  Scoccimarro and A. Zentner
for useful conversations.  This work was supported by  the DOE and the Packard Foundation.

\begin{thebibliography}{52}
\expandafter\ifx\csname natexlab\endcsname\relax\def\natexlab#1{#1}\fi
\expandafter\ifx\csname bibnamefont\endcsname\relax
  \def\bibnamefont#1{#1}\fi
\expandafter\ifx\csname bibfnamefont\endcsname\relax
  \def\bibfnamefont#1{#1}\fi
\expandafter\ifx\csname citenamefont\endcsname\relax
  \def\citenamefont#1{#1}\fi
\expandafter\ifx\csname url\endcsname\relax
  \def\url#1{\texttt{#1}}\fi
\expandafter\ifx\csname urlprefix\endcsname\relax\def\urlprefix{URL }\fi
\providecommand{\bibinfo}[2]{#2}
\providecommand{\eprint}[2][]{\url{#2}}

\bibitem[{\citenamefont{Kamenshchik et~al.}(2001)\citenamefont{Kamenshchik,
  Moschella, and Pasquier}}]{KamMosPas01}
\bibinfo{author}{\bibfnamefont{A.}~\bibnamefont{Kamenshchik}},
  \bibinfo{author}{\bibfnamefont{U.}~\bibnamefont{Moschella}},
  \bibnamefont{and} \bibinfo{author}{\bibfnamefont{V.}~\bibnamefont{Pasquier}},
  \bibinfo{journal}{Phys. Lett. B} \textbf{\bibinfo{volume}{511}},
  \bibinfo{pages}{265} (\bibinfo{year}{2001}), \eprint{gr-qc/0103004}.

\bibitem[{\citenamefont{Bili\'c et~al.}(2002)\citenamefont{Bili\'c, Tupper, and
  Viollier}}]{BilTupVio02}
\bibinfo{author}{\bibfnamefont{N.}~\bibnamefont{Bili\'c}},
  \bibinfo{author}{\bibfnamefont{G.~B.} \bibnamefont{Tupper}},
  \bibnamefont{and} \bibinfo{author}{\bibfnamefont{R.~D.}
  \bibnamefont{Viollier}}, \bibinfo{journal}{Phys. Lett. B}
  \textbf{\bibinfo{volume}{535}}, \bibinfo{pages}{17} (\bibinfo{year}{2002}),
  \eprint{astro-ph/0111325}.

\bibitem[{\citenamefont{Bento et~al.}(2002)\citenamefont{Bento, Bertolami, and
  Sen}}]{BenBerSen02}
\bibinfo{author}{\bibfnamefont{M.~C.} \bibnamefont{Bento}},
  \bibinfo{author}{\bibfnamefont{O.}~\bibnamefont{Bertolami}},
  \bibnamefont{and} \bibinfo{author}{\bibfnamefont{A.~A.} \bibnamefont{Sen}},
  \bibinfo{journal}{\prd} \textbf{\bibinfo{volume}{66}},
  \bibinfo{pages}{043507} (\bibinfo{year}{2002}), \eprint{gr-qc/0202064}.

\bibitem[{\citenamefont{Padmanabhan and Roy~Choudhury}(2002)}]{PadCho02}
\bibinfo{author}{\bibfnamefont{T.}~\bibnamefont{Padmanabhan}} \bibnamefont{and}
  \bibinfo{author}{\bibfnamefont{T.}~\bibnamefont{Roy~Choudhury}},
  \bibinfo{journal}{\prd} \textbf{\bibinfo{volume}{66}},
  \bibinfo{pages}{081301(R)} (\bibinfo{year}{2002}), \eprint{hep-th/0205055}.

\bibitem[{\citenamefont{Sen}(2002)}]{Sen02}
\bibinfo{author}{\bibfnamefont{A.}~\bibnamefont{Sen}} (\bibinfo{year}{2002}),
  \eprint{hep-th/0203265, hep-th/0204143, hep-th/0203211}.

\bibitem[{\citenamefont{Gibbons}(1998)}]{Gib98}
\bibinfo{author}{\bibfnamefont{G.~W.} \bibnamefont{Gibbons}},
  \bibinfo{journal}{Nuc. Phys. B} \textbf{\bibinfo{volume}{514}},
  \bibinfo{pages}{603} (\bibinfo{year}{1998}), \eprint{hep-th/9801106}.

\bibitem[{\citenamefont{Makler et~al.}(2003)\citenamefont{Makler, de~Oliveira,
  and Waga}}]{MakOliWag03}
\bibinfo{author}{\bibfnamefont{M.}~\bibnamefont{Makler}},
  \bibinfo{author}{\bibfnamefont{S.~Q.} \bibnamefont{de~Oliveira}},
  \bibnamefont{and} \bibinfo{author}{\bibfnamefont{I.}~\bibnamefont{Waga}},
  \bibinfo{journal}{Phys. Lett. B} \textbf{\bibinfo{volume}{555}},
  \bibinfo{pages}{1} (\bibinfo{year}{2003}), \eprint{astro-ph/0209486}.

\bibitem[{\citenamefont{Bertolami et~al.}(2004)\citenamefont{Bertolami, Sen,
  Sen, and Silva}}]{BerSenSenSil04}
\bibinfo{author}{\bibfnamefont{O.}~\bibnamefont{Bertolami}},
  \bibinfo{author}{\bibfnamefont{A.~A.} \bibnamefont{Sen}},
  \bibinfo{author}{\bibfnamefont{S.}~\bibnamefont{Sen}}, \bibnamefont{and}
  \bibinfo{author}{\bibfnamefont{P.~T.} \bibnamefont{Silva}},
  \bibinfo{journal}{Mon. Not. R. Atron. Soc.} \textbf{\bibinfo{volume}{353}},
  \bibinfo{pages}{329} (\bibinfo{year}{2004}), \eprint{astro-ph/0402387}.

\bibitem[{\citenamefont{Bento et~al.}(2003)\citenamefont{Bento, Bertolami, and
  Sen}}]{BenBerSen03}
\bibinfo{author}{\bibfnamefont{M.~C.} \bibnamefont{Bento}},
  \bibinfo{author}{\bibfnamefont{O.}~\bibnamefont{Bertolami}},
  \bibnamefont{and} \bibinfo{author}{\bibfnamefont{A.~A.} \bibnamefont{Sen}},
  \bibinfo{journal}{\prd} \textbf{\bibinfo{volume}{67}},
  \bibinfo{pages}{063003} (\bibinfo{year}{2003}), \eprint{astro-ph/0210468}.

\bibitem[{\citenamefont{{Sandvik} et~al.}(2004)\citenamefont{{Sandvik},
  {Tegmark}, {Zaldarriaga}, and {Waga}}}]{SanTegZalWag04}
\bibinfo{author}{\bibfnamefont{H.}~\bibnamefont{{Sandvik}}},
  \bibinfo{author}{\bibfnamefont{M.}~\bibnamefont{{Tegmark}}},
  \bibinfo{author}{\bibfnamefont{M.}~\bibnamefont{{Zaldarriaga}}},
  \bibnamefont{and} \bibinfo{author}{\bibfnamefont{I.}~\bibnamefont{{Waga}}},
  \bibinfo{journal}{\prd} \textbf{\bibinfo{volume}{69}},
  \bibinfo{pages}{123524} (\bibinfo{year}{2004}), \eprint{astro-ph/0212114}.

\bibitem[{\citenamefont{Be\c{c}a et~al.}(2003)\citenamefont{Be\c{c}a, Avelino,
  de~Carvalho, and Martins}}]{BecAveCarMar03}
\bibinfo{author}{\bibfnamefont{L.~M.~G.} \bibnamefont{Be\c{c}a}},
  \bibinfo{author}{\bibfnamefont{P.~P.} \bibnamefont{Avelino}},
  \bibinfo{author}{\bibfnamefont{J.~P.~M.} \bibnamefont{de~Carvalho}},
  \bibnamefont{and} \bibinfo{author}{\bibfnamefont{C.~J. A.~P.}
  \bibnamefont{Martins}}, \bibinfo{journal}{\prd}
  \textbf{\bibinfo{volume}{67}}, \bibinfo{pages}{101301(R)}
  (\bibinfo{year}{2003}), \eprint{astro-ph/0303564}.

\bibitem[{\citenamefont{Avelino et~al.}(2004)\citenamefont{Avelino, Be\c{c}a,
  de~Carvalho, Martins, and Copeland}}]{Aveetal04}
\bibinfo{author}{\bibfnamefont{P.~P.} \bibnamefont{Avelino}},
  \bibinfo{author}{\bibfnamefont{L.~M.~G.} \bibnamefont{Be\c{c}a}},
  \bibinfo{author}{\bibfnamefont{J.~P.~M.} \bibnamefont{de~Carvalho}},
  \bibinfo{author}{\bibfnamefont{C.~J. A.~P.} \bibnamefont{Martins}},
  \bibnamefont{and} \bibinfo{author}{\bibfnamefont{E.~J.}
  \bibnamefont{Copeland}}, \bibinfo{journal}{\prd}
  \textbf{\bibinfo{volume}{69}}, \bibinfo{pages}{041301}
  (\bibinfo{year}{2004}), \eprint{astro-ph/0306493}.

\bibitem[{\citenamefont{Bili\'c et~al.}(2004)\citenamefont{Bili\'c, Lindebaum,
  Tupper, and Viollier}}]{Biletal04}
\bibinfo{author}{\bibfnamefont{N.}~\bibnamefont{Bili\'c}},
  \bibinfo{author}{\bibfnamefont{R.~J.} \bibnamefont{Lindebaum}},
  \bibinfo{author}{\bibfnamefont{G.~B.} \bibnamefont{Tupper}},
  \bibnamefont{and} \bibinfo{author}{\bibfnamefont{R.~D.}
  \bibnamefont{Viollier}}, \bibinfo{journal}{JCAP}
  \textbf{\bibinfo{volume}{11}}, \bibinfo{pages}{008} (\bibinfo{year}{2004}),
  \eprint{astro-ph/0307214}.

\bibitem[{\citenamefont{{Hu} and {Eisenstein}}(1999)}]{HuEis99}
\bibinfo{author}{\bibfnamefont{W.}~\bibnamefont{{Hu}}} \bibnamefont{and}
  \bibinfo{author}{\bibfnamefont{D.~J.} \bibnamefont{{Eisenstein}}},
  \bibinfo{journal}{\prd} \textbf{\bibinfo{volume}{59}},
  \bibinfo{pages}{083509} (\bibinfo{year}{1999}), \eprint{astro-ph/9809368}.

\bibitem[{\citenamefont{{Hu}}(1998)}]{Hu98}
\bibinfo{author}{\bibfnamefont{W.}~\bibnamefont{{Hu}}}, \bibinfo{journal}{\apj}
  \textbf{\bibinfo{volume}{506}}, \bibinfo{pages}{485} (\bibinfo{year}{1998}),
  \eprint{astro-ph/9801234}.

\bibitem[{\citenamefont{Reis et~al.}(2003)\citenamefont{Reis, Waga, Calv\~{a}o,
  and Jor\'as}}]{ReiWagCalJor03}
\bibinfo{author}{\bibfnamefont{R.~R.~R.} \bibnamefont{Reis}},
  \bibinfo{author}{\bibfnamefont{I.}~\bibnamefont{Waga}},
  \bibinfo{author}{\bibfnamefont{M.~O.} \bibnamefont{Calv\~{a}o}},
  \bibnamefont{and} \bibinfo{author}{\bibfnamefont{S.~E.}
  \bibnamefont{Jor\'as}}, \bibinfo{journal}{\prd}
  \textbf{\bibinfo{volume}{68}}, \bibinfo{pages}{061302(R)}
  (\bibinfo{year}{2003}), \eprint{astro-ph/0306004}.

\bibitem[{\citenamefont{Reis et~al.}(2005)\citenamefont{Reis, Makler, and
  Waga}}]{ReiMakWag05}
\bibinfo{author}{\bibfnamefont{R.~R.~R.} \bibnamefont{Reis}},
  \bibinfo{author}{\bibfnamefont{M.}~\bibnamefont{Makler}}, \bibnamefont{and}
  \bibinfo{author}{\bibfnamefont{I.}~\bibnamefont{Waga}},
  \bibinfo{journal}{Class. Quantum Grav.} \textbf{\bibinfo{volume}{22}},
  \bibinfo{pages}{353} (\bibinfo{year}{2005}), \eprint{astro-ph/0501613}.

\bibitem[{\citenamefont{{Hu} et~al.}(2000)\citenamefont{{Hu}, {Barkana}, and
  {Gruzinov}}}]{HuBarGru00}
\bibinfo{author}{\bibfnamefont{W.}~\bibnamefont{{Hu}}},
  \bibinfo{author}{\bibfnamefont{R.}~\bibnamefont{{Barkana}}},
  \bibnamefont{and}
  \bibinfo{author}{\bibfnamefont{A.}~\bibnamefont{{Gruzinov}}},
  \bibinfo{journal}{\prl} \textbf{\bibinfo{volume}{85}}, \bibinfo{pages}{1158}
  (\bibinfo{year}{2000}), \eprint{astro-ph/0003365}.

\bibitem[{\citenamefont{{Scherrer}}(2004)}]{Sch04}
\bibinfo{author}{\bibfnamefont{R.~J.} \bibnamefont{{Scherrer}}},
  \bibinfo{journal}{\prl} \textbf{\bibinfo{volume}{93}},
  \bibinfo{pages}{011301} (\bibinfo{year}{2004}), \eprint{astro-ph/0402316}.

\bibitem[{\citenamefont{{Armendariz-Picon}
  et~al.}(2000)\citenamefont{{Armendariz-Picon}, {Mukhanov}, and
  {Steinhardt}}}]{ArmMukSte00}
\bibinfo{author}{\bibfnamefont{C.}~\bibnamefont{{Armendariz-Picon}}},
  \bibinfo{author}{\bibfnamefont{V.}~\bibnamefont{{Mukhanov}}},
  \bibnamefont{and} \bibinfo{author}{\bibfnamefont{P.~J.}
  \bibnamefont{{Steinhardt}}}, \bibinfo{journal}{\prl}
  \textbf{\bibinfo{volume}{85}}, \bibinfo{pages}{4438} (\bibinfo{year}{2000}),
  \eprint{astro-ph/0004134}.

\bibitem[{\citenamefont{{Arkani-Hamed}
  et~al.}(2004)\citenamefont{{Arkani-Hamed}, {Cheng}, {Luty}, and
  {Mukohyama}}}]{ArkCheLutMuk03}
\bibinfo{author}{\bibfnamefont{N.}~\bibnamefont{{Arkani-Hamed}}},
  \bibinfo{author}{\bibfnamefont{H.-C.} \bibnamefont{{Cheng}}},
  \bibinfo{author}{\bibfnamefont{M.~A.} \bibnamefont{{Luty}}},
  \bibnamefont{and}
  \bibinfo{author}{\bibfnamefont{S.}~\bibnamefont{{Mukohyama}}},
  \bibinfo{journal}{J. High Energy Phys.} \textbf{\bibinfo{volume}{0405}},
  \bibinfo{pages}{074} (\bibinfo{year}{2004}), \eprint{hep-th/0312099}.

\bibitem[{\citenamefont{Chiba et~al.}(2000)\citenamefont{Chiba, Okabe, and
  Yamaguchi}}]{ChiOkaYam00}
\bibinfo{author}{\bibfnamefont{T.}~\bibnamefont{Chiba}},
  \bibinfo{author}{\bibfnamefont{T.}~\bibnamefont{Okabe}}, \bibnamefont{and}
  \bibinfo{author}{\bibfnamefont{M.}~\bibnamefont{Yamaguchi}},
  \bibinfo{journal}{\prd} \textbf{\bibinfo{volume}{62}},
  \bibinfo{pages}{023511} (\bibinfo{year}{2000}), \eprint{astro-ph/9912463}.

\bibitem[{\citenamefont{{Moore} et~al.}(2000)\citenamefont{{Moore}, {Gelato},
  {Jenkins}, {Pearce}, and {Quilis}}}]{Mooetal00}
\bibinfo{author}{\bibfnamefont{B.}~\bibnamefont{{Moore}}},
  \bibinfo{author}{\bibfnamefont{S.}~\bibnamefont{{Gelato}}},
  \bibinfo{author}{\bibfnamefont{A.}~\bibnamefont{{Jenkins}}},
  \bibinfo{author}{\bibfnamefont{F.~R.} \bibnamefont{{Pearce}}},
  \bibnamefont{and} \bibinfo{author}{\bibfnamefont{V.}~\bibnamefont{{Quilis}}},
  \bibinfo{journal}{\apj} \textbf{\bibinfo{volume}{535}}, \bibinfo{pages}{L21}
  (\bibinfo{year}{2000}), \eprint{astro-ph/0002308}.

\bibitem[{\citenamefont{{Yoshida} et~al.}(2000)\citenamefont{{Yoshida},
  {Springel}, and {White}}}]{YosSprWhi00}
\bibinfo{author}{\bibfnamefont{N.}~\bibnamefont{{Yoshida}}},
  \bibinfo{author}{\bibfnamefont{V.}~\bibnamefont{{Springel}}},
  \bibnamefont{and} \bibinfo{author}{\bibfnamefont{S.~D.~M.}
  \bibnamefont{{White}}}, \bibinfo{journal}{\apj}
  \textbf{\bibinfo{volume}{535}}, \bibinfo{pages}{L103} (\bibinfo{year}{2000}),
  \eprint{astro-ph/0002362}.

\bibitem[{\citenamefont{{Garriga} and {Mukhanov}}(1999)}]{GarMuk99}
\bibinfo{author}{\bibfnamefont{J.}~\bibnamefont{{Garriga}}} \bibnamefont{and}
  \bibinfo{author}{\bibfnamefont{V.~F.} \bibnamefont{{Mukhanov}}},
  \bibinfo{journal}{Phys. Lett. B} \textbf{\bibinfo{volume}{458}},
  \bibinfo{pages}{219} (\bibinfo{year}{1999}), \eprint{hep-th/9904176}.

\bibitem[{\citenamefont{{Caldwell} et~al.}(1998)\citenamefont{{Caldwell},
  {Dav\'e}, and {Steinhardt}}}]{CalDavSte98}
\bibinfo{author}{\bibfnamefont{R.~R.} \bibnamefont{{Caldwell}}},
  \bibinfo{author}{\bibfnamefont{R.}~\bibnamefont{{Dav\'e}}}, \bibnamefont{and}
  \bibinfo{author}{\bibfnamefont{P.~J.} \bibnamefont{{Steinhardt}}},
  \bibinfo{journal}{\prl} \textbf{\bibinfo{volume}{80}}, \bibinfo{pages}{1582}
  (\bibinfo{year}{1998}), \eprint{astro-ph/9708069}.

\bibitem[{\citenamefont{{Bean} and {Dore}}(2004)}]{BeaDor04}
\bibinfo{author}{\bibfnamefont{R.}~\bibnamefont{{Bean}}} \bibnamefont{and}
  \bibinfo{author}{\bibfnamefont{O.}~\bibnamefont{{Dore}}},
  \bibinfo{journal}{\prd} \textbf{\bibinfo{volume}{69}},
  \bibinfo{pages}{083503} (\bibinfo{year}{2004}), \eprint{astro-ph/0307100}.

\bibitem[{\citenamefont{{Eisenstein} and {Hu}}(1999)}]{EisHu99}
\bibinfo{author}{\bibfnamefont{D.~J.} \bibnamefont{{Eisenstein}}}
  \bibnamefont{and} \bibinfo{author}{\bibfnamefont{W.}~\bibnamefont{{Hu}}},
  \bibinfo{journal}{\apj} \textbf{\bibinfo{volume}{511}}, \bibinfo{pages}{5}
  (\bibinfo{year}{1999}), \eprint{astro-ph/9710252}.

\bibitem[{\citenamefont{{Gordon} and {Hu}}(2004)}]{GorHu04}
\bibinfo{author}{\bibfnamefont{C.}~\bibnamefont{{Gordon}}} \bibnamefont{and}
  \bibinfo{author}{\bibfnamefont{W.}~\bibnamefont{{Hu}}},
  \bibinfo{journal}{\prd} \textbf{\bibinfo{volume}{70}},
  \bibinfo{pages}{083003} (\bibinfo{year}{2004}), \eprint{astro-ph/0406496}.

\bibitem[{\citenamefont{Press and Schechter}(1974)}]{PreSch74}
\bibinfo{author}{\bibfnamefont{W.~H.} \bibnamefont{Press}} \bibnamefont{and}
  \bibinfo{author}{\bibfnamefont{P.}~\bibnamefont{Schechter}},
  \bibinfo{journal}{\ApJ} \textbf{\bibinfo{volume}{187}}, \bibinfo{pages}{425}
  (\bibinfo{year}{1974}).

\bibitem[{\citenamefont{{Sheth} and {Tormen}}(1999)}]{SheTor99}
\bibinfo{author}{\bibfnamefont{R.~K.} \bibnamefont{{Sheth}}} \bibnamefont{and}
  \bibinfo{author}{\bibfnamefont{B.}~\bibnamefont{{Tormen}}},
  \bibinfo{journal}{\mnras} \textbf{\bibinfo{volume}{308}},
  \bibinfo{pages}{119} (\bibinfo{year}{1999}), \eprint{astro-ph/9901122}.

\bibitem[{\citenamefont{{Barkana} et~al.}(2001)\citenamefont{{Barkana},
  {Haiman}, and {Ostriker}}}]{BarHaiOst01}
\bibinfo{author}{\bibfnamefont{R.}~\bibnamefont{{Barkana}}},
  \bibinfo{author}{\bibfnamefont{Z.}~\bibnamefont{{Haiman}}}, \bibnamefont{and}
  \bibinfo{author}{\bibfnamefont{J.~P.} \bibnamefont{{Ostriker}}},
  \bibinfo{journal}{\apj} \textbf{\bibinfo{volume}{558}}, \bibinfo{pages}{482}
  (\bibinfo{year}{2001}), \eprint{astro-ph/0102304}.

\bibitem[{\citenamefont{{Somerville} et~al.}(2003)\citenamefont{{Somerville},
  {Bullock}, and {Livio}}}]{SomBulLiv03}
\bibinfo{author}{\bibfnamefont{R.~S.} \bibnamefont{{Somerville}}},
  \bibinfo{author}{\bibfnamefont{J.~S.} \bibnamefont{{Bullock}}},
  \bibnamefont{and} \bibinfo{author}{\bibfnamefont{M.}~\bibnamefont{{Livio}}},
  \bibinfo{journal}{\apj} \textbf{\bibinfo{volume}{593}}, \bibinfo{pages}{616}
  (\bibinfo{year}{2003}), \eprint{astro-ph/0303481}.

\bibitem[{\citenamefont{{Yoshida} et~al.}(2003)\citenamefont{{Yoshida}, {Abel},
  {Hernquist}, and {Sugiyama}}}]{YosAbeHerSug03}
\bibinfo{author}{\bibfnamefont{N.}~\bibnamefont{{Yoshida}}},
  \bibinfo{author}{\bibfnamefont{T.}~\bibnamefont{{Abel}}},
  \bibinfo{author}{\bibfnamefont{L.}~\bibnamefont{{Hernquist}}},
  \bibnamefont{and}
  \bibinfo{author}{\bibfnamefont{N.}~\bibnamefont{{Sugiyama}}},
  \bibinfo{journal}{\apj} \textbf{\bibinfo{volume}{592}}, \bibinfo{pages}{645}
  (\bibinfo{year}{2003}), \eprint{astro-ph/0301645}.

\bibitem[{\citenamefont{Narayanan et~al.}(2000)\citenamefont{Narayanan,
  Spergel, Dav\'e, and Ma}}]{NarSpeDavMa00}
\bibinfo{author}{\bibfnamefont{V.~K.} \bibnamefont{Narayanan}},
  \bibinfo{author}{\bibfnamefont{D.~N.} \bibnamefont{Spergel}},
  \bibinfo{author}{\bibfnamefont{R.}~\bibnamefont{Dav\'e}}, \bibnamefont{and}
  \bibinfo{author}{\bibfnamefont{C.-P.} \bibnamefont{Ma}},
  \bibinfo{journal}{\ApJ} \textbf{\bibinfo{volume}{543}}, \bibinfo{pages}{L103}
  (\bibinfo{year}{2000}), \eprint{astro-ph/0005095}.

\bibitem[{\citenamefont{Kogut et~al.}(2003)}]{Kogetal03}
\bibinfo{author}{\bibfnamefont{A.}~\bibnamefont{Kogut}} \bibnamefont{et~al.},
  \bibinfo{journal}{\ApJS} \textbf{\bibinfo{volume}{148}}, \bibinfo{pages}{161}
  (\bibinfo{year}{2003}), \eprint{astro-ph/0302213}.

\bibitem[{\citenamefont{{Kamionkowski} and {Liddle}}(2000)}]{KamLid00}
\bibinfo{author}{\bibfnamefont{M.}~\bibnamefont{{Kamionkowski}}}
  \bibnamefont{and} \bibinfo{author}{\bibfnamefont{A.~R.}
  \bibnamefont{{Liddle}}}, \bibinfo{journal}{\prl}
  \textbf{\bibinfo{volume}{84}}, \bibinfo{pages}{4525} (\bibinfo{year}{2000}),
  \eprint{astro-ph/9911103}.

\bibitem[{\citenamefont{{Zentner} and {Bullock}}(2003)}]{ZenBul03}
\bibinfo{author}{\bibfnamefont{A.}~\bibnamefont{{Zentner}}} \bibnamefont{and}
  \bibinfo{author}{\bibfnamefont{J.~S.} \bibnamefont{{Bullock}}},
  \bibinfo{journal}{\apj} \textbf{\bibinfo{volume}{598}}, \bibinfo{pages}{49}
  (\bibinfo{year}{2003}), \eprint{astro-ph/0304292}.

\bibitem[{\citenamefont{Armendariz-Picon and Lim}(2005)}]{ArmPicLim05}
\bibinfo{author}{\bibfnamefont{C.}~\bibnamefont{Armendariz-Picon}}
  \bibnamefont{and} \bibinfo{author}{\bibfnamefont{E.~A.} \bibnamefont{Lim}}
  (\bibinfo{year}{2005}), \eprint{astro-ph/0505207}.

\bibitem[{\citenamefont{Springel et~al.}(2001)\citenamefont{Springel, Yoshida,
  and White}}]{SprYosWhi01}
\bibinfo{author}{\bibfnamefont{V.}~\bibnamefont{Springel}},
  \bibinfo{author}{\bibfnamefont{N.}~\bibnamefont{Yoshida}}, \bibnamefont{and}
  \bibinfo{author}{\bibfnamefont{S.~D.~M.} \bibnamefont{White}},
  \bibinfo{journal}{New Astronomy} \textbf{\bibinfo{volume}{6}},
  \bibinfo{pages}{79} (\bibinfo{year}{2001}), \eprint{astro-ph/0003162}.

\bibitem[{\citenamefont{Springel and Hernquist}(2002)}]{SprHer02}
\bibinfo{author}{\bibfnamefont{V.}~\bibnamefont{Springel}} \bibnamefont{and}
  \bibinfo{author}{\bibfnamefont{L.}~\bibnamefont{Hernquist}},
  \bibinfo{journal}{Mon. Not. R. Astron. Soc.} \textbf{\bibinfo{volume}{333}},
  \bibinfo{pages}{649} (\bibinfo{year}{2002}), \eprint{astro-ph/0111016}.

\bibitem[{\citenamefont{{Markevich} et~al.}(2004)}]{Maretal04}
\bibinfo{author}{\bibfnamefont{M.}~\bibnamefont{{Markevich}}}
  \bibnamefont{et~al.}, \bibinfo{journal}{\apj} \textbf{\bibinfo{volume}{606}},
  \bibinfo{pages}{819} (\bibinfo{year}{2004}), \eprint{astro-ph/0309303}.

\bibitem[{\citenamefont{{Furlanetto} and {Loeb}}(2002)}]{FurLoe02}
\bibinfo{author}{\bibfnamefont{S.~R.} \bibnamefont{{Furlanetto}}}
  \bibnamefont{and} \bibinfo{author}{\bibfnamefont{A.}~\bibnamefont{{Loeb}}},
  \bibinfo{journal}{\apj} \textbf{\bibinfo{volume}{565}}, \bibinfo{pages}{854}
  (\bibinfo{year}{2002}), \eprint{astro-ph/0107567}.

\bibitem[{\citenamefont{{Gnedin} and {Ostriker}}(2001)}]{GneOst01}
\bibinfo{author}{\bibfnamefont{O.}~\bibnamefont{{Gnedin}}} \bibnamefont{and}
  \bibinfo{author}{\bibfnamefont{J.~P.} \bibnamefont{{Ostriker}}},
  \bibinfo{journal}{\apj} \textbf{\bibinfo{volume}{561}}, \bibinfo{pages}{61}
  (\bibinfo{year}{2001}), \eprint{astro-ph/0010436}.

\bibitem[{\citenamefont{{Miralda-Escud\'e}}(2002)}]{Mir02}
\bibinfo{author}{\bibfnamefont{J.}~\bibnamefont{{Miralda-Escud\'e}}},
  \bibinfo{journal}{\apj} \textbf{\bibinfo{volume}{564}}, \bibinfo{pages}{60}
  (\bibinfo{year}{2002}), \eprint{astro-ph/0002050}.

\bibitem[{\citenamefont{{Natarajan} et~al.}(2002)\citenamefont{{Natarajan},
  {Loeb}, {Kneib}, and {Smail}}}]{NatLoeKneSma02}
\bibinfo{author}{\bibfnamefont{P.}~\bibnamefont{{Natarajan}}},
  \bibinfo{author}{\bibfnamefont{A.}~\bibnamefont{{Loeb}}},
  \bibinfo{author}{\bibfnamefont{J.}~\bibnamefont{{Kneib}}}, \bibnamefont{and}
  \bibinfo{author}{\bibfnamefont{I.}~\bibnamefont{{Smail}}},
  \bibinfo{journal}{\apj Lett.} \textbf{\bibinfo{volume}{580}},
  \bibinfo{pages}{L17} (\bibinfo{year}{2002}), \eprint{astro-ph/0207045}.

\bibitem[{\citenamefont{{Sand} et~al.}(2002)\citenamefont{{Sand}, {Treu}, and
  {Ellis}}}]{SanTreEll02}
\bibinfo{author}{\bibfnamefont{D.~J.} \bibnamefont{{Sand}}},
  \bibinfo{author}{\bibfnamefont{T.}~\bibnamefont{{Treu}}}, \bibnamefont{and}
  \bibinfo{author}{\bibfnamefont{R.~S.} \bibnamefont{{Ellis}}},
  \bibinfo{journal}{\apj} \textbf{\bibinfo{volume}{574}}, \bibinfo{pages}{L129}
  (\bibinfo{year}{2002}), \eprint{astro-ph/0207048}.

\bibitem[{\citenamefont{{Spergel} and {Steinhardt}}(2000)}]{SpeSte00}
\bibinfo{author}{\bibfnamefont{D.~N.} \bibnamefont{{Spergel}}}
  \bibnamefont{and} \bibinfo{author}{\bibfnamefont{P.~J.}
  \bibnamefont{{Steinhardt}}}, \bibinfo{journal}{\prl}
  \textbf{\bibinfo{volume}{84}}, \bibinfo{pages}{3760} (\bibinfo{year}{2000}),
  \eprint{astro-ph/9909386}.

\bibitem[{\citenamefont{{Kodama} and {Sasaki}}(1984)}]{KodSas84}
\bibinfo{author}{\bibfnamefont{H.}~\bibnamefont{{Kodama}}} \bibnamefont{and}
  \bibinfo{author}{\bibfnamefont{M.}~\bibnamefont{{Sasaki}}},
  \bibinfo{journal}{Prog. Theor. Phys. Suppl.} \textbf{\bibinfo{volume}{78}},
  \bibinfo{pages}{1} (\bibinfo{year}{1984}).

\bibitem[{\citenamefont{{Khlopov} et~al.}(1985)\citenamefont{{Khlopov},
  {Malomed}, and {Zeldovich}}}]{KhlMalZel85}
\bibinfo{author}{\bibfnamefont{M.~Y.} \bibnamefont{{Khlopov}}},
  \bibinfo{author}{\bibfnamefont{B.~A.} \bibnamefont{{Malomed}}},
  \bibnamefont{and} \bibinfo{author}{\bibfnamefont{Y.~B.}
  \bibnamefont{{Zeldovich}}}, \bibinfo{journal}{\mnras}
  \textbf{\bibinfo{volume}{215}}, \bibinfo{pages}{575} (\bibinfo{year}{1985}).

\bibitem[{\citenamefont{{Widrow} and {Kaiser}}(1993)}]{WidKai93}
\bibinfo{author}{\bibfnamefont{L.~M.} \bibnamefont{{Widrow}}} \bibnamefont{and}
  \bibinfo{author}{\bibfnamefont{N.}~\bibnamefont{{Kaiser}}},
  \bibinfo{journal}{\apj} \textbf{\bibinfo{volume}{416}}, \bibinfo{pages}{L71}
  (\bibinfo{year}{1993}).

\bibitem[{\citenamefont{{Peebles}}(2000)}]{Pee00}
\bibinfo{author}{\bibfnamefont{P.~J.~E.} \bibnamefont{{Peebles}}},
  \bibinfo{journal}{\apj Lett.} \textbf{\bibinfo{volume}{534}},
  \bibinfo{pages}{L127} (\bibinfo{year}{2000}), \eprint{astro-ph/0002495}.

\end{thebibliography}

\end{document}